\documentstyle[sprocl]{article}

\def\be{\begin{equation}}
\def\ee{\end{equation}}
\def\bea{\begin{eqnarray}}
\def\eea{\end{eqnarray}}
\newcommand{\moresim}{\raisebox{-0.5ex}{\mbox{ $\stackrel{>}{\sim}$ }} }

\begin{document}

\title{COLOR AND SPIN IN QUARKONIUM PRODUCTION\footnote
	{Invited talk presented at 
	Fifth International Workshop on Hard Probes in Nuclear Collisions, 
	Lisbon, September 1997, and at XXVII International Symposium on 
	Multiparticle Dynamics, Frascati, September 1997.}}

\author{ERIC BRAATEN}

\address{Department of Physics, Ohio State University\\
	 Columbus, OH 43210, USA\\
	 E-mail: braaten@pacific.mps.ohio-state.edu} 


\maketitle
\abstracts{
I describe the NRQCD factorization approach to the inclusive production 
of heavy quarkonium, contrasting it with the color-singlet and 
color evaporation models.
These approaches differ dramatically in their assumptions 
about the roles played by color and spin in the production process.
They also differ dramatically in their predictions for the production
of charmonium at large transverse momentum.}

\section{Introduction}
The production of heavy quarkonium in high energy collisions 
is important, because there are several quarkonium states with 
clean experimental signatures through their decays into lepton pairs.  
Measurements of the production rate of these states can be used to
test our understanding both of heavy-quark production in high energy 
collisions and of the formation of bound states from heavy quark pairs.  
The NRQCD factorization approach, which is based on the use 
of an effective field theory called nonrelativistic QCD
to exploit the large mass of the heavy quark, 
has led to dramatic progress in our understanding of 
inclusive quarkonium production. 
Below, I describe the basic ideas underlying the NRQCD
factorization approach, emphasizing its implications for
the role played by color and spin in inclusive production.

\section{Simple Production Models}

Before 1995, most of the effort to understand the inclusive production 
of charmonium was carried out within either the
{\it color-singlet model} or the {\it color evaporation model}.
Both of these models have roots
that go back to 1975 shortly after the discovery of charmonium.  
These two models make diametrically opposite assumptions 
about the roles played by color and
spin in the production process. 

\subsection{Color-singlet Model}

In the color-singlet model, charmonium states are interpreted as 
nonrelativistic bound states of a $c$ and $\bar c$ interacting
through a confining potential.
The $J/\psi$, for example, is identified as a bound state of a
$c \bar c$ pair in a color-singlet $^3S_1$ state.  I denote such a state by
$c \bar c_1 (^3S_1)$, where the subscript specifies the color state (1 for
singlet, 8 for octet) and the argument specifies the angular-momentum quantum
numbers.  The $\chi_{cJ}$ is identified as a $c \bar c_1 (^3P_J)$ bound state.  
The only role of gluons in the color-singlet model is to generate the 
potential that binds the $c$ and $\bar c$.

The color-singlet model provides a prescription for calculating not only the
inclusive production rate for a quarkonium state, but also its inclusive decay
rate into light hadrons and its decay rate into leptons or photons.
For example, the formula for the decay rate
of the $J/\psi$ into light hadrons in the color-singlet model is
\begin{equation}
\Gamma (J/\psi) \;=\;
\widehat \Gamma \big( c \bar c_1 (^3S_1) \big)
        {|R_{J/\psi} (0)|^2 \over 4 \pi} ,
\label{Gam-CSM}
\end{equation}
where $\widehat \Gamma$ is proportional to the annihilation rate at
threshold of a $c \bar c$ pair in the state $c \bar c_1(^3S_1)$.
The last factor in (\ref{Gam-CSM})
is the square of the wavefunction at the origin, which gives the probability
for the $c \bar c$ pair in the $J/\psi$ to be close enough to annihilate.  

In the color-singlet model, it is assumed that a $c \bar c$ 
pair that is produced in a high energy collision 
will bind to form a given charmonium state
only if the  $c \bar c$ pair is created in a color-singlet state with
angular-momentum quantum numbers that match those of the bound state.  
For example, to form a $J/\psi$, the $c \bar c$ pair must be produced 
in a color-singlet $^3S_1$ state.
The formula for the inclusive production cross
section for the $J/\psi$ in the color-singlet model is
\begin{equation}
\sigma (J/\psi) \;=\;
\widehat \sigma \big( c \bar c_1 ( ^3S_1) \big)
        {| R_{J/\psi}(0)|^2 \over 4 \pi} ,
\label{sig-CSM}
\end{equation}
where $\widehat \sigma$ is proportional to the production rate for a $c \bar c$
pair at threshold in the state $c \bar c_1 (^3S_1)$.
 The last factor in (\ref{sig-CSM})
is the probability that a pointlike $c \bar c $ pair in the state $c
\bar c_1 (^3S_1)$ will bind to form a $J/\psi$, which is again given by the
square of the wavefunction at the origin.

The color-singlet model is an extremely economical phenomenonological
framework for calculating quarkonium production.  The factors $\hat \Gamma$
and $\hat \sigma$ in (1) and (2) can be computed using QCD perturbation theory
in terms of the charm quark mass $m_c$ and the running coupling constant
$\alpha_s(m_c)$.  The only other phenomenological parameters are a single
wavefunction factor for each spin multiplet:  $R_{J/\psi}(0)$ for the S-wave
states $J/\psi$ and $\eta_c$, $R^\prime_\chi (0)$ for the P-wave states
$\chi_{c0}$, $\chi_{c1}$, $\chi_{c2}$, and $h_c$, etc.  
Moreover, these wavefunction
factors can be determined from experimental measurements of the decay rates.
Thus, the color-singlet model predicts the inclusive production cross sections
for all quarkonium states in terms of a single phenomenological parameter
$m_c$.  

The color-singlet model gives definite predictions
for the dependence of the cross section on the polarization of
the charmonium state.   It also predicts that ratios 
$\sigma (H) / \sigma (J/\psi)$ of the
cross sections of quarkonium states with different $J^{PC}$
quantum numbers should vary dramatically from
process to process, because of angular momentum selection rules.
Unfortunately, these predictions have proved to be
wrong.  Dramatic variations in the ratios $\sigma (H)/ \sigma (J/\psi)$ have
not been seen experimentally and there is little if any evidence for 
spin asymmetries in the cross sections.

The most dramatic failure of the color-singlet model came in 1995, when the
CDF collaboration measured the production cross sections for charmonium states
at the Tevatron $p \bar p$ collider.  
They used a vertex detector to separate prompt
production of charmonium from production through the decay of $b$ quarks.
They found that the cross sections for the direct production of $J/\psi$ and
$\psi^\prime$ at large transverse momentum were larger than the predictions
of the color-singlet model by about a factor of 30.\cite{Sansoni}  
This dramatic discrepancy
marked the final demise of the color-singlet model.  May it rest in peace.

\subsection{Color Evaporation Model}

In the color evaporation model, the probability that a $c \bar c$ 
pair produced in a high energy collision will bind to form a 
specific charmonum state is assumed to be almost completely
independent of the color and spin state of the $c \bar c$ pair. 
The basic assumption is that the
exchange and emission of soft gluons destroys any correlations between the
color and spin state of the $c \bar c$ pair when it is created and the
quantum numbers of the final $c \bar c$ bound state. 
In the color evaporation model, the formula for the inclusive cross 
section for producing a $J/\psi$ is
conventionally written in the form
\begin{equation}
\sigma (J/\psi) \;=\;
\widehat \sigma (c \bar c : 4m_c^2 < s < 4 m^2_D) \; f_{J/\psi} ,
\end{equation}
where $\widehat \sigma$ is the cross section 
for producing a $c \bar c$ pair with
invariant mass below the $D \bar D$ threshold.  This cross section is 
summed over all the color and spin states of the $c \bar c$ pair.  
The factor $f_{J/\psi}$ in (3) is
a phenomenological parameter that gives the fraction of $c \bar c$ pairs
below the $D \bar D$ threshold that form a $J/\psi$.

The color evaporation model is much less ambitious than the color-singlet
model.  It makes no attempt to relate production cross sections to 
annihilation decays.  However it is still a fairly economical 
phenomenological framework for calculating
quarkonium production.  The  cross section $\widehat \sigma$ in
(3) can be calculated using perturbative QCD as a function of $m_c, m_D$, and
$\alpha_s(m_c)$.  The choice of $4 m_D^2$ as the upper endpoint for the
integration over $s$ is completely arbitrary, so we should regard
$s_{\rm max} = 4 m_D^2$ as a phenomenological parameter.  The remaining
phenomenological parameters are a fraction $f_H$ for each quarkonium state
$H$.  Thus the color evaporation model gives predictions for inclusive
production cross sections in all high energy processes in terms of $m_c$,
$s_{\rm max}$, and a fraction $f_H$ for each quarkonium state.

The color evaporation model gives simple predictions for the ratios
of cross sections for different quarkonium states
and for the dependence of the cross sections on the
polarization of the quarkonium state.
It predicts that ratios of cross sections 
are independent of the production process:
\begin{equation}
{\sigma (H) \over \sigma (J/\psi)} \;=\; {f_H \over f_{J/\psi}} .
\end{equation}
The model also  predicts that cross sections 
are independent of the polarization of
the charmonium state.  For example, the fraction of $J/\psi$'s that are
transversely polarized is predicted to be 2/3 for any production process.
Thus the color evaporation model can be ruled out by experimental 
measurements of a nonzero spin asymmetry for the $J/\psi$ in any process  
or of a variation of the $\chi_c$-to-$J/\psi$ ratio in different processes.

The color evaporation model is not yet dead.  It is basically a
phenomenological model, so it can only be killed by experimental data.  
I believe that its demise is only a matter of time until sufficiently 
accurate experimental data is available.  The model is based on
far too simplistic a picture of the effects of soft gluons 
on color and spin in quarkonium production.

\section{The NRQCD Factorization Approach}

The {\it NRQCD factorization approach} to quarkonium production 
was developed by Bodwin, Braaten, and Lepage in 1995.\cite{B-B-L}
As in the color evaporation model, a $c \bar c$ pair in any color and 
spin state has a nonzero probability of binding to form a given 
charmonium state.  As in the color-singlet model, that probability 
depends strongly on the color and angular-momentum state.
Thus the roles of spin and color in this approach are somewhere
intermediate between their roles in the color-singlet and 
color evaporation models.
However, in contrast to these models, the NRQCD factorization approach
is based firmly on QCD.

\subsection{Quarkonium as a Nonrelativistic Bound State}

The NRQCD factorization approach to annihilation decays and inclusive
production is based on the fact that heavy quarkonium is a
nonrelativistic bound state.  Its properties are determined by QCD in terms of
essentially only 2 parameters:  the QCD coupling constant $\alpha_s$ and the
heavy quark mass $M$.  However, because it is a nonrelativistic bound state,
there are a number of different energy scales that play an important role in
quarkonium physics.  These scales include
\begin{itemize}

\item 
$M$, the mass of the heavy quark,
 which sets the scale for the mass of the bound state,

\item 
$M v$, the typical momentum of the heavy quark in quarkonium, which
sets the scale for the size of a quarkonium state,

\item 
$M v^2$, the typical kinetic energy of the heavy quark, which is also
the scale for splittings between radial excitations and between orbital
angular momentum excitations,

\item 
$\Lambda_{\rm QCD}$, the scale of nonperturbative effects associated
with light quarks and gluons.

\end{itemize}
The scales $Mv$ and $M v^2$ owe their existence to the fact that a small
parameter $v$ is generated dynamically in a nonrelativistic bound state by the
balance between kinetic energy and potential energy.  If the interactions
between the quark and antiquark are described by a potential $V(R)$, the
balance can be expressed in the form
\begin{equation}
Mv^2 \;\sim\; V(R)  \qquad {\rm for} \; R \;\sim\; 1/(Mv).
\end{equation}
If the mass $M$ is large enough, the potential is essentially Coulombic:
$V(R) \sim \alpha_s (R) / R$.  The condition (5) then reduces to
$v \sim \alpha_s(M v)$.  If the mass $M$ is not so large, the potential may
be dominated by the linear confinement region:  $V(R) \sim \kappa^2R$.  In this
case, the condition (5) gives $v \sim (\kappa/M)^{2/3}$.  In either case, we
get a dynamically-generated small parameter $v$.

In Table \ref{tab:scales}, 
I list the energy scales $M$, $Mv$, and $Mv^2$ for charmonium and
bottomonium and also what they would be for toponium if the top quark did not
decay too rapidly to form bound states.  For the scale $M$,
I have taken half the mass of the lowest quarkonium state.  
For the scale $Mv^2$, I have taken the average of the
energy splittings for the first radial excitation and the first orbital
angular momentum excitation.  The scale listed for $Mv$ is just the geometric
mean of $M$ and $Mv^2$.  By dividing $Mv^2$ by $M$, we find that $v^2$ is
about 1/3 for charmonium, 1/10 for bottomonium, and about 1/100 for toponium.
While 1/3 is not tiny, it is small enough that we can hope to use it as an
expansion parameter for relativistic corrections.

\begin{table}[t]
\caption{Momentum scales for quarkonium systems}
\label{tab:scales}
\vspace{0.2cm}
\begin{center}
\footnotesize
\begin{tabular}{lccc}
\hline
       & $c \bar c$ & $b \bar b$ & $t \bar t$ \\
\hline
$M$    &  1.5 GeV   &  4.7 GeV   &  180 GeV   \\
$Mv$   &  0.9 GeV   &  1.5 GeV   &   16 GeV   \\
$Mv^2$ &  0.5 GeV   &  0.5 GeV   &  1.5 GeV   \\
\hline
\end{tabular}
\end{center}
\end{table}

In Table \ref{tab:alphas}, 
I list the running coupling constant of QCD at each of the
momentum scales for charmonium, bottomonium, and toponium.  For charmonium and
for bottomonium, the scale $Mv^2$ is definitely in the strong-coupling
region.  At the scale $M$, $\alpha_s$ is small enough that we should be able
to use perturbation theory to calculate the effects of that scale.

\subsection{Nonrelativistic QCD}

The large mass of the charm and bottom quarks gives us two small numbers
$\alpha_s (M)$ and $v^2$ that we can exploit to understand heavy quarkonium
physics.  In order to exploit the smallness of $\alpha_s(M)$, we need
to solve the problem of separating the scale $M$ from the smaller momentum
scales $Mv$, $Mv^2$, and $\Lambda_{\rm QCD}$ that involve nonperturbative
physics.  In order to exploit the smallness of $v^2$,
we need some way of organizing nonperturbative
effects according to how they scale with $v$.  These
problems were solved by Peter Lepage and collaborators\cite{L-M-N-M-H} 
by introducing an effective field
theory called {\it nonrelativistic QCD} (NRQCD).

NRQCD is a formulation of QCD in which heavy quarks are treated as
nonrelativistic particles.  Instead of describing the heavy quark and
antiquark by a single 4-component Dirac field $\Psi$, 
they are described by separate 2-component Pauli fields $\psi$ 
and $\chi^\dagger$.  The lagrangian for NRQCD has the form 
\begin{equation}
{\cal L}_{\rm NRQCD} \;=\; 
{\cal L}_{\rm light} \;+\; {\cal L}_0 \;+\; {\cal L}_2 \;+\; \cdots,
\end{equation}
where ${\cal L}_{\rm light}$ is the usual QCD lagrangian for
the gluons and light quarks and antiquarks, ${\cal L}_0$ is the minimal NRQCD
lagrangian,
\begin{equation}
{\cal L}_0 \;=\; 
\psi^\dagger \left( i D_0 + {{\bf D}^2 \over 2M} \right) \psi 
+ \chi^\dagger \Big(i D_0 - {{\bf D}^2 \over 2M} \Big) \chi,
\end{equation}
and ${\cal L}_2$, ${\cal L}_4$, $\ldots$ are a series of ``improvement terms''
that can be added to make NRQCD reproduce the physics of full QCD to higher
and higher order in $v^2$.  The minimal QCD lagrangian ${\cal L}_0$ is
invariant under heavy-quark spin symmetry, which mixes heavy quarks with spin
up and spin down.  The $v^2$-improvement term ${\cal L}_2$ includes 
four terms:
a $({\bf D}^2)^2$ term that includes the relativistic correction to the
kinetic energy, a ${\bf D} \cdot {\bf E}$
term, and ${\bf D} \times {\bf E} \cdot \mbox{\boldmath{$\sigma$}}$ and
${\bf B} \cdot \mbox{\boldmath{$\sigma$}}$ terms that break the spin
symmetry.  The coefficients of these four terms depend only on the momentum
scale $M$ and they can therefore be calculated as perturbation series in
$\alpha_s(M)$.

\begin{table}[t]
\caption{The QCD coupling constant at the momentum
scales for quarkonium systems}
\label{tab:alphas}
\vspace{0.2cm}
\begin{center}
\footnotesize
\begin{tabular}{lccc}
\hline
 		  & $c \bar c$ & $b \bar b$ & $t \bar t$ \\
\hline
$\alpha_s(M)$     & 0.35       & 0.22       & 0.11       \\
$\alpha_s(M v)$   & 0.52       & 0.35       & 0.16       \\
$\alpha_s(M v^2)$ & $\moresim$ 1    & $\moresim$ 1    & 0.35       \\
\hline
\end{tabular}
\end{center}
\end{table}

The NRQCD collaboration has calculated 
the spectrum for bottomonium and charmonium using Monte Carlo
simulations of lattice NRQCD.\cite{NRQCD:spectrum} 
These calculations demonstrate convincingly that 
NRQCD provides an effective framework for describing
heavy quarkonium physics.
Using the lagrangian ${\cal L}_{\rm light} + {\cal L}_0$, 
the splittings between spin-averaged energy levels are reproduced
to an accuracy of about 30\% for charmonium and 10\% for bottomonium, 
which are consistent with errors of relative order $v^2$.  Because this
lagrangian has spin symmetry, it gives no spin splittings.  After adding the
$v^2$-improvement term ${\cal L}_2$, the accuracy of the spin-averaged
splittings is improved to 10\% for charmonium and 1\% for bottomonium,
consistent with errors of order $v^4$.
This term also gives the correct spin splittings up to errors of order $v^2$. 
Higher accuracy could presumably be achieved by 
adding the $v^4$-improvement term ${\cal L}_4$.

NRQCD provides a simple solution to the problem of separating the scales $M$
from the lower momentum scales $Mv$, $Mv^2$, and $\Lambda_{\rm QCD}$.  The
improvement terms in the NRQCD lagrangian have coefficients that must be tuned
to reproduce the physics of heavy quarks in full QCD.  These terms compensate
for the fact that NRQCD does not treat correctly the physics of relativistic
heavy quarks whose momenta are of order $M$.  Their coefficients therefore
depend only on the momentum scale $M$.  Since they depend only on the
physics at length scales of order $1/M$, we call them
``short-distance coefficients.''  In NRQCD, all effects of the
scale $M$ are taken into account through the coefficients of the improvement
terms in the NRQCD lagrangian.  Thus the NRQCD lagrangian itself
provides us with the desired separation of scales.
The effects of the scale $M$ are all contained in the 
short-distance coefficients.

NRQCD also provides a solution to the problem of organizing 
nonperturbative effects in such a way as to exploit the smallness of
$v^2$.\cite{Braaten}  
It leads to a simple picture of the structure of a charmonium state,
which is described most easily in terms of the Fock state expansion in 
Coulomb gauge.  This gauge makes the dynamics of a 
nonrelativistic bound state particularly transparent, and it 
allows for a sensible Fock state expansion
since there are no negative norm states.  For example,
the Fock state decomposition for the $J/\psi$ in Coulomb gauge
has the form
\begin{equation}
| J/\psi \rangle \;=\; 
\sum_{c \bar c} \psi^{J/\psi}_{c \bar c_1 (^3S_1)} 
	| c \bar c_1 (^3S_1) \rangle
\;+\; \sum_{c \bar c g} \psi^{J/\psi}_{c \bar c_8 (^3P_J) g} 
	| c \bar c_8 (^3P_J) + g \rangle
+ \cdots .
\end{equation}
The probability of each Fock state scales in a definite way with $v$.
Only the $c \bar c_1 (^3S_1)$ Fock state has a probability of order 1.
All the higher Fock states have probabilities  suppressed by powers of $v$.
The higher Fock state with the greatest probability is 
$c \bar c_8 (^3P_J) + g$, whose probability is of order $v^2$.  
The next most important Fock state is $c \bar c_8 (^1S_0) + g$, 
whose probability is of order $v^3$.  
One of the next most important Fock states is $c \bar c_8 (^3S_1) + g$, 
whose probability is of order $v^4$.  
By exploiting this Fock space structure,
one can derive velocity-scaling rules
that determine how matrix elements of local operators
in a quarkonium state scale with $v$.
If nonperturbative effects can be organized into matrix elements,
their relative magnitudes can be estimated using the velocity-scaling rules.

\subsection{NRQCD Factorization Formula for Decays}

The NRQCD factorization approach to annihilation decays involves separating
the scale $M$ from the lower momentum scales $Mv$, $Mv^2$, and 
$\Lambda_{\rm QCD}$, and then exploiting the fact that quarkonium physics 
involves two small numbers, $\alpha_s(M)$ and $v^2$.  
The smallness of $\alpha_s(M)$ is exploited by calculating the
effects of the scale $M$ as perturbation series in $\alpha_s(M)$.  
The smallness of $v^2$ is exploited by using the velocity-scaling rules to
estimate the magnitudes of nonperturbative matrix elements involving scales
of order $Mv$ and smaller.

We proceed to outline the derivations of the factorization formula for
annihilation decay rates.  The inclusive rate for the decay of a 
quarkonium state $H$ into light hadrons can be expressed in the 
following schematic form:
\begin{equation}
\Gamma(H) \;=\; 
	{1 \over 2 M_H} \; \sum_{Q \bar Q} \sum_{\rm hadrons}
\left | \psi^H_{Q \bar Q} \otimes T_{Q \bar Q \to {\rm hadrons}} \right |^2 ,
\label{Gam-H}
\end{equation}
where $\psi^H_{Q \bar Q}$ is the wavefunction for $H$ to consist of a $Q$
and $\bar Q$ and $T_{Q \bar Q \to {\rm hadrons}}$ is the T-matrix element for
producing a particular final state consisting of light hadrons.  
The sum over all final-state hadrons in (\ref{Gam-H}) can be replaced by a
sum over all final-state partons.  This simply amounts to a change of 
basis for color-singlet final states.
The final-state partons can be separated into hard partons 
with momenta of order $M$
and soft partons with much smaller momenta.  The resulting expression
for the decay rate has the form
\begin{equation}
\Gamma(H) \;=\; 
	{1 \over 2M_H} \; \sum_{Q \bar Q} \; \sum_{\rm hard} \; \sum_{\rm soft}
\left| \psi^H_{Q \bar Q} \otimes 
	T_{Q \bar Q \to {\rm hard + soft}} \right|^2 .
\end{equation}
The soft gluons in the final state can be emitted from the $Q$ or
$\bar Q$, from the hard partons, or from the
soft partons in the final state.  The T-matrix element
also involves virtual soft gluons that connect the initial $Q$ and $\bar Q$,
the final hard partons, and the final soft partons.  Standard factorization
methods of perturbative QCD can be used to show that for every pair of hard
partons whose 4-momenta are not collinear, there is a cancellation between the
corrections from virtual soft gluons exchanged between the two partons and the
interference terms between real soft gluons emitted by the two partons.  A
similar cancellation occurs between the corrections from
virtual soft gluons exchanged between 
a hard parton and the $Q \bar Q$ pair and the interference terms 
from real soft gluons emitted by the hard parton and the $Q \bar Q$ pair.  
The only soft gluons that survive after these
cancellations are the virtual gluons exchanged between the $Q$ and $\bar Q$
and the real gluons emitted from the $Q$ and $\bar Q$.  But the effects of
these gluons can be absorbed into the wavefunction 
$\psi^H_{Q \bar Q + {\rm soft}}$ for $H$ to consist 
of a $Q \bar Q$ pair plus soft gluons.
The resulting expression for the decay rate has the form
\begin{equation}
\Gamma (H) \;=\; 
	{1 \over 2 M_H} \; \sum_{Q \bar Q} \; 
	\sum_{\rm hard} \; \sum_{\rm soft} 
\left| \psi^H_{Q \bar Q + {\rm soft}} \otimes 
	T_{Q \bar Q \to {\rm hard}} \right |^2.
\end{equation}
The sum over $Q \bar Q$ states includes a sum over their color (1 or 8) and
angular-momentum ($^{2S+1}L_J$) quantum numbers, which we denote
collectively by $n$, and an integral over the magnitude of their relative
momentum $q$.  Because of the wavefunction factor, 
the integral over $q$ has support only for $q$ of order $Mv$.
All dependence on  the scale $Mv$ can be removed from the
T-matrix element by expanding $T$ in powers of $q$.  
Denoting the expansion coefficients by $\widehat T$
and ssociating the factors of $q$ and the integral
over $q$ with the wavefunction factor, we obtain
\begin{equation}
\Gamma (H) \;=\; 
	{1 \over 2 M_H} \; \sum_n \left ( \sum_{\rm hard}
\left | \widehat T_{Q \bar Q [n] \to {\rm hard}} \right |^2 \right ) 
\left ( \int_q q^L \; \sum_{\rm soft} \; 
\left | \psi^H_{Q \bar Q [n] + {\rm soft}} \right |^2 \right) .
\label{Gam-sep}
\end{equation}
The sum over $n$
includes all color and angular-momentum states of a $Q \bar Q$ pair.  
Each term in the sum is the product of a short-distance
factor that involves only the scale $M$ and a long-distance factor that
involves only scales of order $Mv$ or smaller.  The short-distance factor is
proportional to the annihilation rate at threshold for a $Q \bar Q$ pair in
the state $n$.  The long-distance factor
measures the probability of finding the $Q \bar Q$ pair at the
same point in the state $n$ in the quarkonium state $H$.

The factorization formula (\ref{Gam-sep}) is simply a statement about the
separation of scales in QCD, so it does not necessarily require NRQCD.  
However NRQCD provides a convenient prescription for carrying out this
separation of scales.  The long-distance factor in (\ref{Gam-sep}) can be
expressed as the expectation value in the 
quarkonium state $H$ of a local gauge-invariant
operator ${\cal O}_n$ in NRQCD.  The short-distance factor is proportional 
to the imaginary part of the coefficient of the
operator ${\cal O}_n$ in the NRQCD lagrangian.
Thus (\ref{Gam-sep}) can be written
\begin{eqnarray}
\Gamma (H) \;=\; {1 \over 2M_H} \; \sum_n \; 
\widehat \Gamma_{Q \bar Q[n]} \langle H | {\cal O}_n | H \rangle .
\label{Gam-fact}
\end{eqnarray}
This is the NRQCD factorization formula for the decay rate.  
Since the coefficient $\widehat \Gamma$ 
involves only the scale $M$,
it can be calculated as a perturbation series in $\alpha_s(M)$.
The matrix elements $\langle H | {\cal O}_n | H \rangle$ 
can in principle be calculated nonperturbatively using Monte Carlo 
simulations of lattice NRQCD.\cite{B-S-K}
Thus the NRQCD factorization formula can be used to calculate
the decay rate from first principles.

The NRQCD factorization formula (\ref{Gam-fact}) contains infinitely 
many terms.  The utility of the formula relies on the fact that NRQCD 
also provides a way of exploiting the smallness of $v^2$.
Each of the matrix elements scales with a definite power of $v$.
In the case of $J/\psi$, the largest matrix element is
$\langle J/\psi | {\cal O}_1(^3S_1) | J/\psi \rangle$, which
is proportional to the square of the wavefunction at the origin 
and scales like $v^3$.  The next most important matrix elements
are the expectation values of ${\cal O}_8(^1S_0)$, which scales like $v^6$,
and of ${\cal O}_8(^3S_1)$ and ${\cal O}_8(^3P_J)$, which scale like $v^7$.
The relative magnitudes of the terms in the factorization formula
(\ref{Gam-fact}) are determined by the order in $v$ of the matrix elements 
and by the order in $\alpha_s$ of the short-distance coefficients.

\subsection{NRQCD Factorization Formula for Production}

The NRQCD factorization approach to quarkonium production 
is based on the fact that a sufficiently inclusive cross section
satisfies a factorization formula analogous to (\ref{Gam-sep}).
All effects involving momentum scales of order $M$ and larger are 
contained in short-distance factors that involve
the T-matrix element for producing a $Q \bar Q$ pair in the state $n$
with given 4-momentum $P$.  All effects involving
momentum scales of order $M v$ and smaller are contained in
long-distance factors that involve the amplitude for a $Q \bar Q$ pair 
in the state $n$ to form the quarkonium state $H$ plus soft partons.
The effective field theory NRQCD provides a convenient prescription 
for carrying out this separation of scales.
It can also be used to exploit the small number $v^2$ by
using velocity-scaling rules to determine how the 
long-distance factors scale with $v$.

The factorization formula for the
differential cross section for producing a 
quarkonium state $H$ with 4-momentum $P$ has the form
\begin{equation}
d \sigma \big( H(P) \big) \;=\; \sum_n d \widehat \sigma_{Q \bar Q [n, P]}
\langle {\cal O}_n^H \rangle,
\label{dsig-fact}
\end{equation}
where the sum over $n$ includes all color and angular-momentum states of the
$Q \bar Q$ pair.  The short-distance factor $\widehat \sigma$ is proportional
to the cross section for producing a $Q \bar Q$ pair at threshold in the state
$n$ with total 4-momentum $P$.  
Since they involve only momentum scales of order $M$ and larger, 
the short-distance factors can be calculated as perturbation series in
$\alpha_s(M)$.  The long-distance factor 
$\langle {\cal O}_n^H \rangle$, which can be expressed as an
NRQCD matrix element, 
is proportional to the probability that a $Q \bar Q$ pair produced at
a point in the state $n$ will form the quarkonium state $H$ 
plus soft hadrons, whose energies
in the $H$ rest frame are of order $M v$ or smaller.  
The derivation of the factorization formula breaks down if the 
4-momentum of the quarkonium is collinear with that of any hadrons 
in the initial state.  It therefore  does
not apply in the diffractive region.  

The NRQCD factorization formula (\ref{Gam-fact}) contains infinitely 
many terms.  The relative importance of these terms depends on several 
factors.  The relative magnitudes of the matrix elements
depends on how they scale with $v$.  The relative magnitudes of the 
short-distance coefficients depends on their order in $\alpha_s$,
and also on how they scale with dimensionless ratios of 
kinematic variables, such as $m_c/p_T$ in the case of production 
at large transverse momentum.  In practice, there are several matrix 
elements of phenomenological importance 
for any given quarkonium state.
Unfortunately we have no effective nonperturbative methods for
calculating these matrix elements, with a few exceptions.  The exceptions are
matrix elements like $\langle {\cal O}_1^{J/\psi}(^3S_1) \rangle$ that can
be related to the wavefunction of the dominant $c \bar c_1$ Fock state.  
The remaining matrix elements must be treated as phenomenological parameters.  
The predictive power of the factorization formula resides in the fact that 
the same matrix elements must describe inclusive quarkonium production 
in all high energy processes.

The factorization formula (\ref{Gam-fact}) is simply a statement
about the separation of scales in QCD.  One can ignore the prediction of 
NRQCD for the relative magnitudes of the matrix elements,
and simply take the formula as a model-independent framework
for analyzing quarkonium production.  Any model that is consistent 
with QCD at short distances must be expressible in this form 
for some choice of the NRQCD matrix elements.  In particular,
the color-singlet and color evaporation models can be expressed 
as specific assumptions about the NRQCD matrix elements.
The color-singlet model reduces to the assumption that the only 
matrix element that is nonzero is the color-singlet matrix element
whose angular-momentum quantum numbers correspond
to those of the $c \bar c$ pair in the dominant Fock state.  
This matrix element is 
$\langle {\cal O}_1^{J/\psi}(^3S_1) \rangle$ for the $J/\psi$
and $\langle {\cal O}_1^{\chi_{cJ}}(^3P_J) \rangle$ for the $\chi_{cJ}$.
The color evaporation model corresponds to the assumption that
the S-wave matrix elements dominate and that they are equal up to 
simple color and spin factors:
\begin{equation}
\langle {\cal O}_1^H(^3S_1) \rangle \;=\;
\langle {\cal O}_1^H(^1S_0) \rangle
\;=\;
6 \; \langle {\cal O}_8^H(^3S_1) \rangle \;=\;
6 \; \langle {\cal O}_8^H(^1S_0) \rangle .
\label{matel-CEM}
\end{equation}
The P-wave matrix elements are all suppressed by a factor
of $(m_D^2 - m_c^2)/m_c^2$, which is of order $v^2$.
NRQCD predicts a much more intricate pattern for the relative 
magnitudes of the matrix elements and this pattern depends on the $J^{PC}$
quantum numbers of the quarkonium state $H$. 
The matrix elements that scale with the leading power of $v$ 
include both the color-singlet model matrix element and 
one of the color evaporation model matrix elements in (\ref{matel-CEM}).
For the $J/\psi$, the only leading matrix element is
$\langle {\cal O}_1^{J/\psi}(^3S_1) \rangle$, which scales like $v^3$. 
For the $\chi_{cJ}$, the leading matrix elements are
$\langle {\cal O}_1^{\chi_{cJ}}(^3P_J) \rangle$
and $\langle {\cal O}_8^{\chi_{cJ}}(^3S_1) \rangle$,
both of which scale like $v^5$.

As a phenomenological framework for calculating quarkonium production,
the NRQCD factorization approach
is much less economical than the color evaporation model.
For each quarkonium state, there are several matrix elements 
of phenomenological importance.  For $J/\psi$ production, 
the largest matrix element is 
$\langle {\cal O}_1^{J/\psi}(^3S_1) \rangle$, which scales like $v^3$. 
Its value can be determined from decays of the $J/\psi$.
The next most important matrix elements are 
$\langle {\cal O}_8^{J/\psi}(^1S_0) \rangle$, 
$\langle {\cal O}_8^{J/\psi}(^3S_1) \rangle$,
and $\langle {\cal O}_8^{J/\psi}(^3P_J) \rangle$, which scale like
$v^6$, $v^7$, and $v^7$, respectively.  These three color-octet matrix elements 
must be treated as independent phenomenological parameters,
in contrast with the single phenomenological parameter $f_{J/\psi}$
in the color evaporation model.

The NRQCD factorization approach differs dramatically from 
the color evaporation model in its predictions for the 
dependence of the cross section on the polarization of the quarkonium state. 
The color evaporation model predicts no dependence on the polarization.
The NRQCD factorization approach predicts a nontrivial dependence, 
because the NRQCD matrix 
elements $\langle {\cal O}_n^H \rangle$ depend on the polarization
of the quarkonium state $H$.

\section{Applications to Charmonium Production}
\label{sec:Apps}

The NRQCD factorization approach has been applied to
charmonium production in almost all possible high energy processes.
Some of these applications are described 
in two recent reviews.\cite{B-F-Y,Beneke}  
I will focus on one specific application for which 
the predictions of the NRQCD factorization approach differ 
significantly from 
both the color-singlet and the color evaporation models.

\subsection{$J/\psi$ at Large Transverse Momentum}
\label{sec:Prompt}

The NRQCD factorization formula leads to a
dramatic prediction for the production of prompt charmonium
at large transverse momentum in $p \bar p$ collisions. 
The prediction is that at sufficiently large $p_T$,
most of the $J/\psi$'s will be transversely polarized.
This prediction follows from several simple steps,
which I proceed to discuss in detail.

The first step in the argument involves the separation of the scale $p_T$
from the lower momentum scales of order $m_c$ and smaller. 
According to standard factorization theorems of 
perturbative QCD,\cite{Collins-Soper}
the inclusive production cross section for
any hadron at large $p_T$ is dominated by {\it fragmentation},
which means that it is produced by the hadronization of a single
high-$p_T$ parton.  As pointed out by Braaten and Yuan,\cite{Braaten-Yuan} 
this factorization theorem applies to 
charmonium states as well as to light hadrons provided that 
$p_T \gg m_c$.\cite{Braaten-Yuan}  In the case of the 
$J/\psi$, the factorization formula is
\begin{equation}
d \sigma (p \bar p \to J/\psi(P) + X) \;=\;
\sum_i \int_0^1 dz
d \sigma (p \bar p \to i(P/z) + X) \;
        D_{i \to J/\psi}(z) \,,
\label{dsig-frag}
\end{equation}
where $d \sigma(p \bar p \to i + X)$ is the inclusive cross section 
for producing a parton of type $i$ with larger transverse momentum
$p_T/z$.  This cross section can be calculated in terms of the 
parton distributions for $p$ and $\bar p$ and parton cross sections 
$d \widehat{\sigma}$ that involve only the scale $p_T/z$ and
can therefore be calculated as power series in $\alpha_s(p_T/z)$.  
The fragmentation function 
$D_{i \to J/\psi}$ gives the probability
that the hadronization of the parton $i$ produces a $J/\psi$ carrying
a fraction $z$ of the parton momentum.
All effects of momentum scales smaller than $p_T$, including 
the effects of the charm quark mass and
nonperturbative effects involved in the formation of the bound state
$J/\psi$, are contained within the fragmentation functions.
The largest fragmentation probabilities are those for
$c$, $\bar c$, and the gluon.  The sum over partons $i$ in (\ref{dsig-frag})
is completely dominate by the gluon, because the cross section 
for producing gluons is so much larger than that for 
producing charm quarks.
The corrections to the factorization formula (\ref{dsig-frag})
from nonfragmentation processes
are suppressed asymptotically by powers of $m_c^2/p_T^2$.
Thus the factorization of the scale $p_T$ leads to the conclusion
that $J/\psi$ production at large $p_T$ is dominated by gluon
fragmentation.

The second step in the argument involves the separation of the scale $m_c$
from the lower momentum scales of order $m_c v$ and smaller.
The NRQCD factorization formula (\ref{dsig-fact}) can be used to express 
the gluon fragmentation function in the form
\begin{equation}
D_{g \to J/\psi}(z) \;=\;
\sum_{n} \widehat d_{g \to c \bar c[n]}(z) \;
	\langle {\cal O}_n^{J/\psi} \rangle \,,
\label{frag-fact}
\end{equation}
where $\widehat d_{g \to c \bar c[n]}$ is the fragmentation function
for producing a $c \bar c$ in the state $n$ carrying a fraction $z$ 
of the momentum of the gluon.  This fragmentation function involves
only momenta of order $m_c$, and it can therefore be calculated as a 
power series in $\alpha_s(m_c)$.  All nonperturbative effects involving the 
binding of the $c \bar c$ pair to form a $J/\psi$ are 
factored into the NRQCD matrix elements.  

According to the color-singlet model, the gluon fragmentation function
(\ref{frag-fact}) should be dominated by the $c \bar c_1(^3S_1)$ term.
The leading contribution to the short-distance coefficient for this term 
is of order $\alpha_s^3$
and comes from the parton process $g^* \to c \bar c g g$.
Keeping only this term in the fragmentation function (\ref{frag-fact}),
the cross section predicted by (\ref{dsig-frag}) is about a factor of
30 below recent CDF data on prompt $J/\psi$ production at the Tevatron.
It is this result that spelled the final demise of the color-singlet model.

According to the color evaporation model, the gluon fragmentation function
(\ref{frag-fact}) should be dominated by the S-wave matrix element
with the largest short-distance coefficient.
We will see that the NRQCD factorization approach leads to the 
same conclusion.  As first pointed out by
Braaten and Fleming,\cite{Braaten-Fleming}, 
the term that dominates is the $c \bar c_8(^3S_1)$
term, which corresponds to the formation of a $J/\psi$ from a 
$c \bar c$ pair that is created in a color-octet $^3S_1$ state.
At leading order in $\alpha_s$, this term in the fragmentation function is
\begin{equation}
D_{g \to J/\psi}(z) \approx
{\pi \alpha_s(m_c) \over 96 m_c^4} \delta(1-z)
\langle {\cal O}_8^{J/\psi}(^3S_1) \rangle .
\label{D-psi}
\end{equation}
The relative importance of the various terms in the fragmentation function 
(\ref{frag-fact}) depends on several different factors.
We will compare each of these factors  for the $c \bar c_1(^3S_1)$ term
and the $c \bar c_8(^3S_1)$ term.
The magnitude of the NRQCD matrix element is determined by its order in
$v$, which is $v^3$ for $\langle {\cal O}^{J/\psi}_1(^3S_1) \rangle$
and $v^7$ for $\langle {\cal O}^{J/\psi}_8(^3S_1) \rangle$.
The magnitude of the short-distance coefficient $\widehat d$ is determined by 
its order in $\alpha_s$.  The coefficient of the 
$c \bar c_1(^3S_1)$ term comes from 
the parton process $g^* \to c \bar c gg$ and
is of order $\alpha_s^3$, while the $c \bar c_1(^3S_1)$ term
has a much larger coefficient of order $\alpha_s$ coming from 
the parton process $g^* \to c \bar c$.
However the relative importance of the terms also depends on the shape 
of the $z$-distribution of $\widehat d$.  The reason is that in the 
expression (\ref{dsig-frag}) for the cross section,
the fragmentation function is folded with the cross section for producing 
a gluon, which is a rapidly falling function of the transverse momentum
$p_T/z$ of the gluon.  Since $d \sigma/dp_T^2$ falls like the inverse 
fourth power of the gluon's transverse momentum,
the cross section is effectively weighted by $z^4$.
This dramatically suppresses the $c \bar c_1(^3S_1)$ term,
which has a rather soft $z$ distribution with $\langle z \rangle \approx 0.4$.
In contrast, there is very little suppression of the $c \bar c_8(^3S_1)$ term
in (\ref{D-psi}), since it is sharply peaked near $z=1$.
Putting all the factors together, the $c \bar c_1(^3S_1)$ term
scales like  $\langle z^4 \rangle \alpha_s^3 v^3$ while the 
$c \bar c_8(^3S_1)$ term scales like  $\alpha_s v^7$.
The suppression of the color-singlet term by $\langle z^4 \rangle \alpha_s^2$
overwhelms the suppression of the color-octet term by $v^4$.
Thus the factorization of the scale $m_c$ leads to the conclusion that
the gluon fragmentation function is dominated by the
$c \bar c_8(^3S_1)$ term.  This term is singled out because it's
short-distance coefficient is the lowest order in $\alpha_s$
and the most sharply peaked near $z=1$.

Unfortunately, we cannot use the formula (\ref{D-psi}) to predict
the cross section at large $p_T$, because the matrix element
$\langle {\cal O}^{J/\psi}_8(^3S_1) \rangle$ has not yet been determined
accurately from other production processes.
However, the shape of the cross section $d \sigma/dp_T$
obtained by inserting (\ref{D-psi}) into (\ref{frag-fact}) is in agreement
with the Tevatron data at large $p_T$.  The value of the matrix element 
can therefore be determined by fitting the data.
The resulting value of $\langle {\cal O}^{J/\psi}_8(^3S_1) \rangle$ 
is consistent with suppression by a factor of $v^4$ relative to the 
color-singlet model matrix element
$\langle {\cal O}_1^{J/\psi}(^3S_1) \rangle$.

The color evaporation model gives a prediction for the $J/\psi$ cross
section at large $p_T$ that is similar to that of the   
NRQCD factorization approach, because the color evaporation model
matrix elements in (\ref{matel-CEM}) include
$\langle {\cal O}^{J/\psi}_8(^3S_1) \rangle$.
However the NRQCD factorization approach differs dramatically
in its predictions for the
dependence of the cross section on the polarization of the $J/\psi$.
The color evaporation model predicts the absence of any spin
asymmetry. As pointed out by Cho and Wise \cite{Cho-Wise}, 
the NRQCD factorization approach predicts that 
$J/\psi$'s should be predominantly
transversely polarized at sufficiently large $p_T$.
The leading term (\ref{D-psi}) in the gluon fragmentation 
function contributes only to the production of $J/\psi$'s 
that are  transversely polarized.
The radiative corrections to the fragmentation function 
were examined by Beneke and Rothstein \cite{Beneke-Rothstein}, 
and they concluded that the spin alignment
at large $p_T$ remains greater than 90\%.
The largest corrections to the spin alignment at values of $p_T$
that have been measured at the Tevatron come from
nonfragmentation contributions that are suppressed by $m_c^2/p_T^2$
relative to the terms in (\ref{dsig-frag}). 
These contributions have been calculated by  
Beneke and Kraemer and by Leibovich \cite{Beneke-Kramer}.
Their prediction is that the spin allignment should 
be almost zero at small $p_T$, that it should begin to turn on
at a $p_T$ of 3 to 5 GeV, and that it should reach
60\% to 80\% of its maximum possible value for $p_T$ of 20 GeV.
An experimental measurement of the spin alignment
in agreement with these predictions would constitute a dramatic triumph
for the NRQCD factorization approach.
It would also spell the final demise of the color evaporation model.

\section{Conclusions}

The NRQCD factorization approach
has led to dramatic progress in our understanding of
inclusive quarkonium production.
Within this approach, the roles played by color and spin 
in the production process are somewhere intermediate between the extreme 
assumptions of the color-singlet and color evaporation models.
In some cases, the NRQCD factorization approach gives
predictions that differ dramatically from those of both the
color-singlet model and the color evaporation model.
A particularly important example is the production 
of $J/\psi$ at large $p_T$ in $p \bar p$ colliders,
where a measurement of the spin asymmetry will provide
a stringent test of our present understanding of inclusive 
quarkonium production.

\section*{Acknowledgments}
This work was supported in part by the U.S.
Department of Energy, Division of High Energy Physics, under
Grant DE-FG02-91-ER40690.

\end{document}